\documentclass[preprint,showpacs,aps]{revtex4}
\usepackage{graphicx,subfigure}
\begin{document}
\title{Two-dimensional percolation with multiple seeds}
\author{Hongting Yang}
\affiliation{School of Science, Wuhan University of Technology, Wuhan 430070, P.R. China}
\author{Stephan Haas}
\affiliation{Department of Physics and Astronomy, University of Southern California, Los Angeles, CA 90089-0484}

\begin{abstract}
We study non-uniform percolation in a two-dimensional cluster growth model with multiple seeds. With increasing
concentration of seeds, the percolation threshold is found to increase monotonically, while the exponents for correlation
length, order parameter, and average cluster size, keep invariant. The scaling law for an infinite square lattice keeps
working for any nonzero concentration of seeds. Abnormal finite-size scaling behaviours happen at low concentration of
seeds.
\end{abstract}

\pacs{64.60.ah,05.10.Ln,05.70.Jk} \maketitle
\section{introduction}
In an ordinary two-dimensional percolation model, the sites or bonds of a lattice are usually distributed uniformly,
whether the lattice is square, triangular, diamond, or in other forms~\citep{stauffer}. However, in nature, the non-uniform
distribution may be more popular than the uniform one. For example, electron density is used to describe the non-uniform
spatial distribution of an electron in materials~\citep{hehre2003}, cancer cells begins in some tissues or organs of the
body, but not the whole body. In view of the wide influence of percolation
theory~\citep{sahimi1994,hunt2009,achlio2009,lai2010,sergey2010,costa2010,riordan2011}, it is essential to study the
nonuniform percolation models and their percolation properties.

In the past decades, an important development towards nonuniform percolation is the study on the correlated percolation,
examples of which are bootstrap percolation~\citep{adler1991,paolo2004,balogh2009,sausset2010}, jamming
percolation~\citep{fisher2006,toninelli2008,jeng2008,ghosh2014}, and directed
percolation~\citep{hinrichsen2000,odor2004,hinrichsen2006,henkel2008,zhou2012,lipowski2012,landes2012,wang2012} etc. Our
model is certainly some kind of correlated percolation model. However, our model originates from a completely different
idea. In our model, clusters start to grow from a number of preoccupied sites. Our model is at first a cluster growth
model. Except one special case, the model displays similar properties as the ordinary non-restricted percolation model. The
only difference is the specific values of percolation thresholds. The critical properties of the model are obtained in the
same way that already used in an ordinary percolation model.

For the convenience, we summarize the formulae here, detailed description of the method can be found
elsewhere~\citep{stauffer}. An interesting quantity of a percolation model is the correlation length defined as
\begin{equation}
  \xi^2=\frac{2\sum_sR_s^2s^2n_s}{\sum_ss^2n_s},
\end{equation}
where $n_s$ is the average number of $s$-clusters per lattice site, $2R_s^2=\sum_{ij}|{\bf r}_i-{\bf r}_j|^2/s^2$ is the
average squared distance between two cluster sites. It is expected to behave as
\begin{equation}
  \xi(p)\sim|p-p_c|^{-\nu}.
\end{equation}
Given the value of $p_c$, one can get the exponent $\nu$ by fitting the values of $\xi(p)$ around $p_c$. However, this
method may give rise to large uncertainties of $\nu$, since the exponent $\nu$ is sensitive to the data points around
$p_c$. On the other hand, the fitted value of $\nu$ for $p>p_c$ may be different from that for $p<p_c$. To circumvent these
uncertainties, one can choose other methods instead. An effective recipe is to introduce the probability $\Pi(p,L)$, which
is the probability that a lattice of linear dimension $L$ percolates at concentration $p$. In an infinite system, $\Pi=1$
above and $\Pi=0$ below $p_c$. As for percolation transitions, only the value of $p_c$ is not enough, we have to introduce
a number of other observables. On a square lattice with periodic boundary conditions, the quantity ${\rm d}\Pi/{\rm d}p$
gives the probability per interval ${\rm d}p$ at concentration $p$, that a wrapping cluster appears for the first time. The
average concentration $p_{av}$, at which, a wrapping cluster appears for the first time is defined as
\begin{equation}
  p_{av}=\int p\left(\frac{{\rm d}\Pi}{{\rm d}p}\right){\rm d}p.
\end{equation}
If we define the width $\Delta$ of the transition region as
\begin{equation}
  \Delta^2=\int (p-p_{av})^2\left(\frac{{\rm d}\Pi}{{\rm d}p}\right){\rm d}p,
\end{equation}
then $\Delta$ can be related to $p_{av}$ via
\begin{equation}
  p_{av}-p_c \propto\Delta.
\end{equation}
In this way, one can first get the value of $p_c$ by fitting the observed thresholds $p_{av}$ and the observed widths
$\Delta$, without prior knowledge of the correlation length exponent $\nu$. With the value of $p_c$ at hand, one can
further obtain the value of $\nu$ through
\begin{equation}
  |p_{av}-p_c|\propto L^{-1/\nu}.
\end{equation}
If an observable $X$ is predicted to scale as $|p-p_c|^{-\lambda}$ in an infinite lattice, then we expect it to obey the
general scaling law
\begin{equation}
  X(L,p)=(p-p_c)^{-\lambda}\tilde{X}\left((p-p_c)L^{1/\nu}\right),
\end{equation}
where $\tilde{X}$ is a scale-independent function. Other two interesting observables are the order parameter, $P_\infty$,
the probability of an occupied site belongs to the infinite cluster, and the average cluster size, $S$. At $p=p_c$, they
behave respectively as
\begin{equation}
  P_\infty(L, p_c)\propto L^{-\beta/\nu},
\end{equation}
\begin{equation}
  S(L, p_c)\propto L^{\gamma/\nu}.
\end{equation}
Here, the value of $P_\infty$ (or $S$) at $p=p_c$ is estimated by linear interpolation between two values of $P_\infty$ (or
$S$) right above and below $p_c$. The exponents for two-dimensional lattice are expected to obey the well-known scaling law
\begin{equation} \label{scalinglaw}
  \gamma+2\beta=2\nu.
\end{equation}
These exponents will be calculated in our model.

\section{model}
Our model can be easily built up by first occupying a number of sites randomly chosen from a two-dimensional square lattice
with $N=L^2$ sites, then one by one, occupy the empty sites being neighbour to the previous occupied sites. After occupying
a qualified empty site (being neighbour to at least one of the previous occupied sites), the list of qualified empty sites
is refreshed by adding the new empty sites being neighbour to the site that was occupied just now. Groups of neighbour
sites form clusters. Thus clusters grow from the multiple seeds, the sites occupied at the very beginning. Since the later
occupied sites in our model are surrounding the previously occupied sites, there could be some differences between our
model and other models. We use $\eta$ to denote the concentration of seeds. For the Eden
model~\citep{eden1961,jullien1985}, $\eta=1/N$, which approaches zero with increasing $L$. When $\eta\ge\tilde{p}_c$, with
$\tilde{p}_c$ the percolation threshold of the usual two-dimensional site percolation~\citep{feng2008,lee2008,yang2012},
the critical region of percolation transition will be covered by the occupying process of the sites chosen as seeds, the
following cluster growth process is therefore less meaningful. Especially, $\eta=1$ is the usual percolation model. So, it
is of interest only in the region with $0<\eta<\tilde{p}_c$.

\begin{figure}[htbp]
\centering
\includegraphics[width=0.5\textwidth]{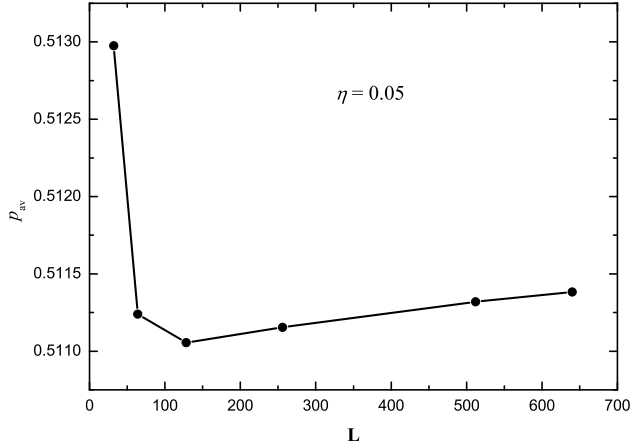}
\caption{\label{f0}The average threshold $p_{av}$ for $\eta=0.05$ changes with the linear dimension $L$. The abnormal
change of $p_{av}$ happens at $L_a=64$ for the selected data points.}
\end{figure}

We have calculated the thresholds $p_{av}$ for many sets of $\eta$ and $L$. For moderate value of $\eta$, the values of
$p_{av}$ decrease with decreasing $L$ as what could be found in a general percolation model. However, an abnormal
phenomenon could be observed for small values of $\eta$. The values of $p_{av}$ first decrease with decreasing $L$, after
passing some critical lattice dimension $L_a$, the $p_{av}$ values increase abnormally with decreasing $L$. Obviously, the
scaling behaviours of $p_{av}$ above and below $L_a$ are different. To get any convergent observable of an infinite system,
one has to choose lattices with $L$ greater than the critical lattice dimension $L_a$. If $L_a$ is too large to meet the
requirement of a computer memory, then no critical information of an infinite system can be obtained. The abnormal change
of $p_{av}$ happens at $L_a=32, 64, 128$ for $\eta=0.1, 0.05, 0.025$ respectively. As an example, the circumstance of
$\eta=0.05$ is shown in FIG.~\ref{f0}.
Simple calculation gives $L_a\propto\eta^{-1}$. Clearly, $L_a\to\infty$ when $\eta\to 0$. The Eden model $(\eta\to 0)$ is
purely a cluster growth model, which is inappropriate to be regarded as a percolation model. Along another trend of
changing $\eta$, there should be a critical value of $\eta$, only below which, the abnormal scaling behaviour of $p_{av}$
happens. This value of $\eta$ for a finite lattice could be figured out by changing the number of seeds one by one. This is
clearly not a easy work. Our model, except the case $\eta=0$, is some kind of constrained percolation model.

\section{results}
\begin{figure}[h]
\centering \subfigure[\ $\eta=0.2$]{\includegraphics[width=0.23\textwidth]{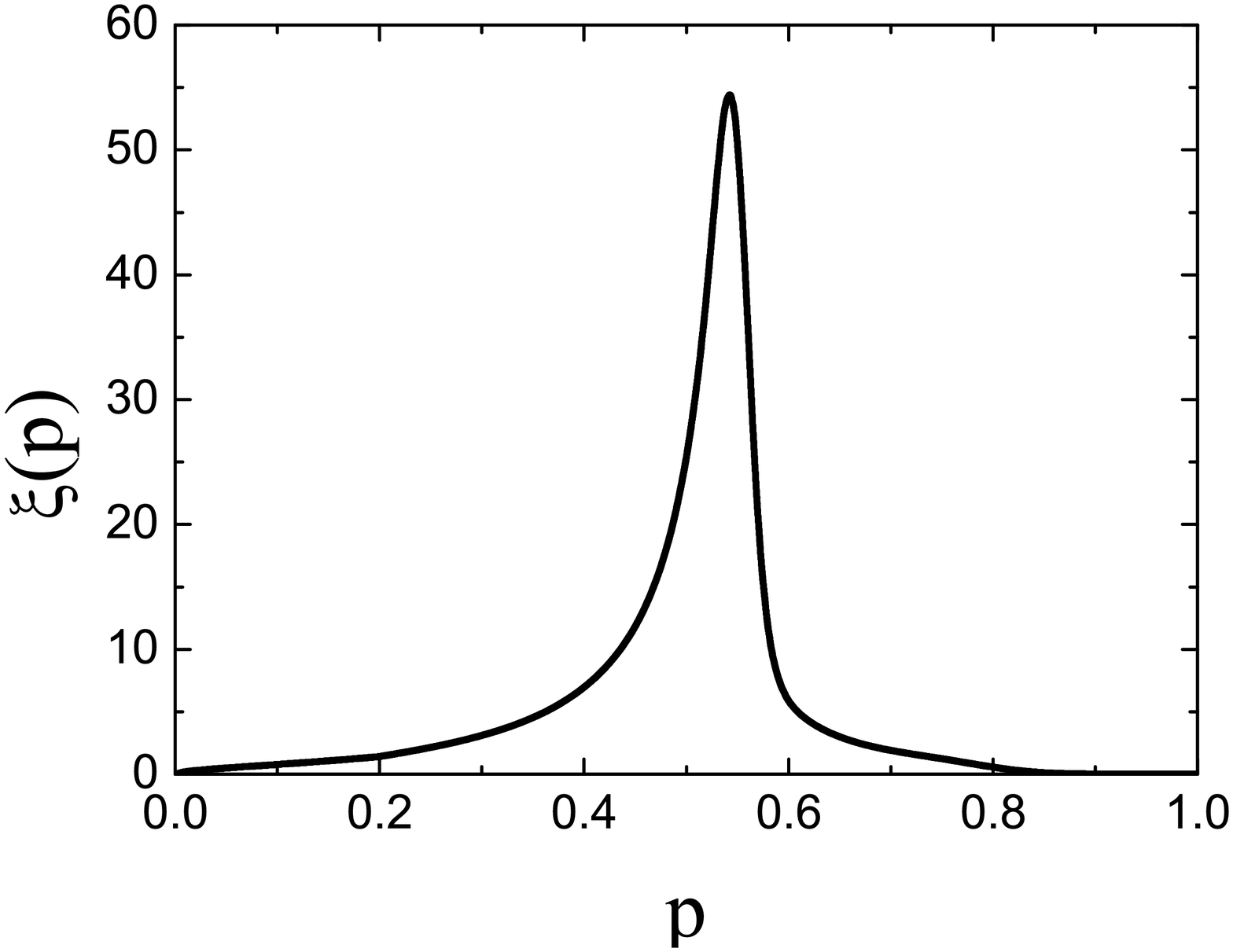}} \subfigure[\
$\eta=0.5$]{\includegraphics[width=0.23\textwidth]{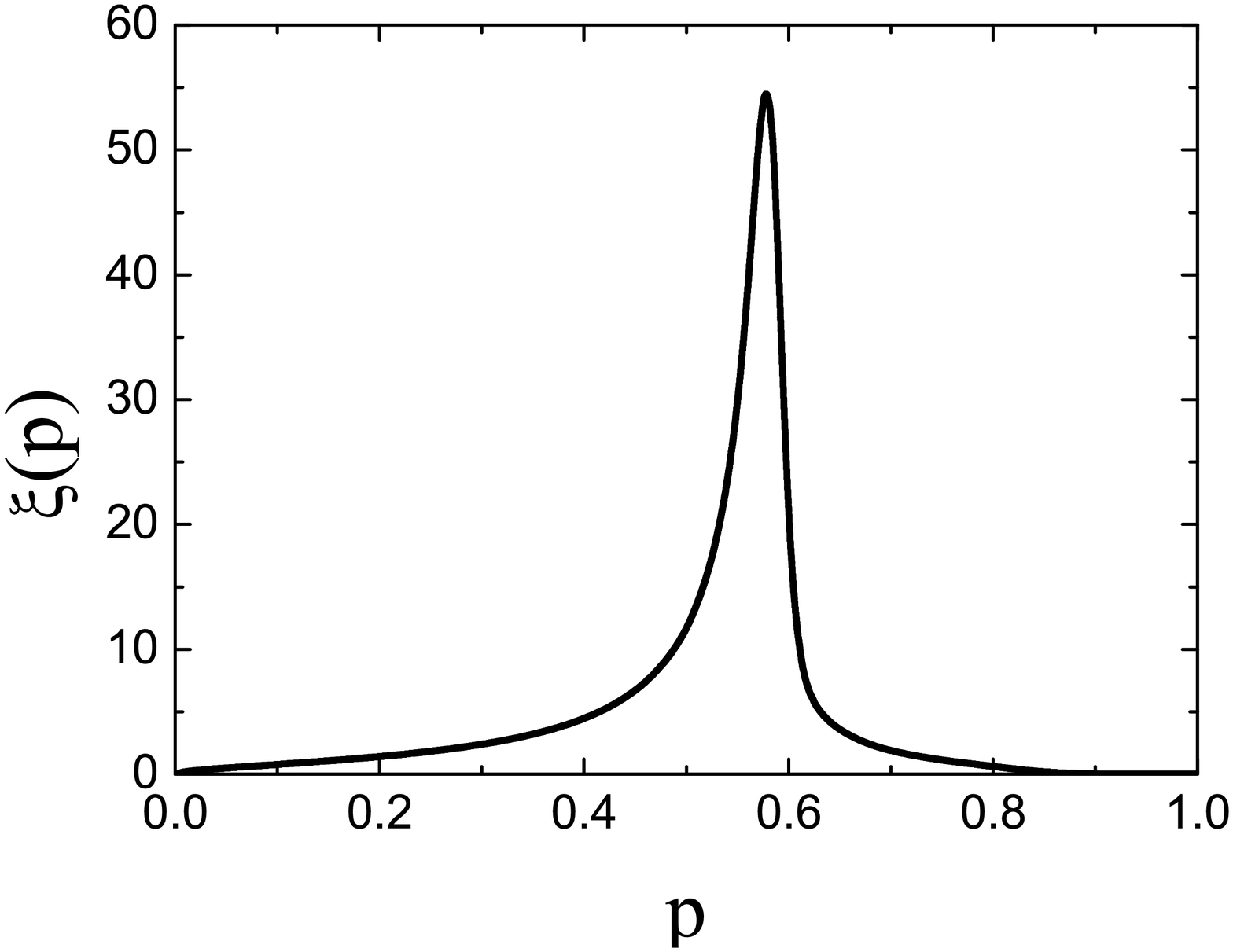}} \caption{The $p$-dependence of $\xi$ for the periodic square
lattice with $L=128$. (a) is for $\eta=0.2$ and (b) for $\eta=0.5$.\label{f1}}
\end{figure}

On a lattice with $N$ sites, there are $N\eta$ seeds, from which clusters start to grow until the lattice is fully occupied
in each run or configuration. Relative to other observables, the computation of $\xi$ needs rather longer CPU time. Given
one value of $\eta$ on the lattice with $L=128$, a common desktop PC with CPU clock speed 2.6 GHz should keep running for
about 14 hours to output the values of $\xi(p)$ averaged on $20000$ runs, and the corresponding $p$-dependence of $\xi(p)$
for $\eta=0.2$ and $\eta=0.5$ is shown in FIG.~\ref{f1}.
The peak value of $\xi$ is insensitive to the value of $\eta$ as it should be. This peak value is larger than that in the
free-boundary lattice with the same $L$, because touching-boundary clusters could become bigger by including the sites
across the periodic boundaries, and a spanning cluster not counted in calculating $\xi$ should be summed if it does not
form a wrapping cluster when the free-boundary condition is switched to the periodic-boundary condition~\citep{yang2012}.
It is worth of noting that, the peak value of $\xi$ appears in different point, one is around 0.54, the other is around
0.58, which implies that percolation thresholds for models with different concentration of seeds are different.

\begin{figure}[htbp]
\centering
\includegraphics[width=0.5\textwidth]{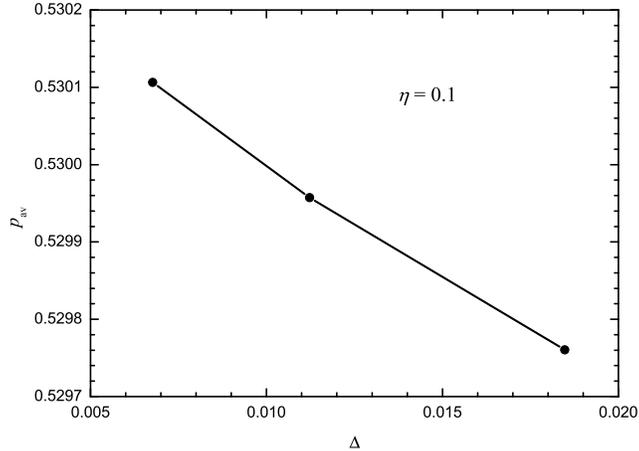}
\caption{\label{f2}The average threshold $p_{av}$ versus the transition width $\Delta$. The concentration of seeds is
$\eta=0.1$, and the data points are respectively for $L$=128, 256, and 512.}
\end{figure}

\begin{figure}[htbp]
\centering
\includegraphics[width=0.5\textwidth]{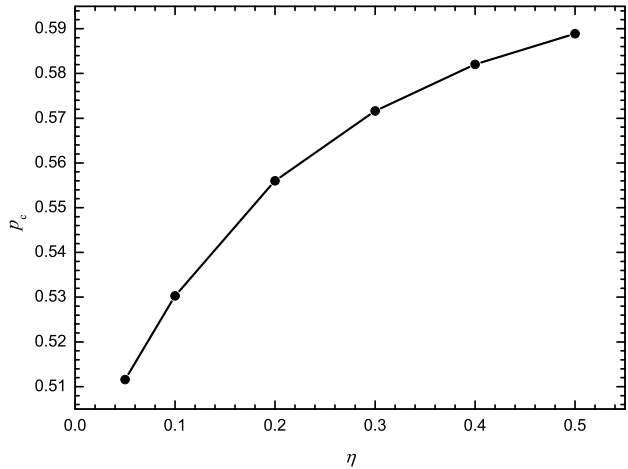}
\caption{\label{f3}The dependence of the percolation thresholds $p_c$ on the concentration of seeds $\eta$.}
\end{figure}

For each set of parameters $L$ and $\eta$, we have calculated $p_{av}$, $\Delta$, $P_\infty$, and $S$. The number of runs
is in the range of $1.3\times10^5$ to $2.3\times10^8$, and the corresponding computation time is 13--16.5 hours. In
FIG.~\ref{f2}, at the concentration of seeds $\eta=0.1$, the average threshold $p_{av}$ versus the width of transition
region $\Delta$ is given for $L$=512, 256, and 128, respectively. Linear fitting gives the percolation threshold for an
infinite system $p_c=0.530298(22)$ for the given $\eta$. The datum for $L$=64 is not adopted in linear fitting since it is
close to the critical lattice dimension 32 for $\eta=0.1$. For $\eta=0.2$, we also choose the data with $L$=512, 256, and
128, to do linear fitting. For $\eta\ge0.2$, we choose the data with $L$=64, 128, and 256 in linear fitting, while choosing
the data with $L$= 256, 512, and 640, for $\eta=0.05$. In the same way, the $p_c$ values for other $\eta$ values
($\eta$=0.5, 0.4, 0.3, 0.2, 0.05) are obtained and summarized in FIG.~\ref{f3}. Obviously, the $p_c$ values decrease with
decreasing $\eta$ values. In other words, if the number of seeds for cluster growing is smaller, the percolation phase
transition will happen earlier. This result seems a little bit strange but is understandable. In our model, the following
occupied sites are gathered to the clusters centered with these seeds occupied at the very beginning. Given more seeds,
which means there are more clusters growing from these seeds to randomly distribute all the occupied sites, thus the
largest cluster in this case will be smaller, and the wrapping cluster will be certainly delayed to appear. Data fitting
gives $p_c=a+be^{c\eta}$, with $a=0.59951(89)$, $b=-0.11056(67)$, and $c=-4.64(11)$. The minimum $p_c=0.4890(16)$ in the
limit $\eta\to0$ is unreachable since it is not a percolation model in this case.

\begin{figure}[htbp]
\centering
\includegraphics[width=0.5\textwidth]{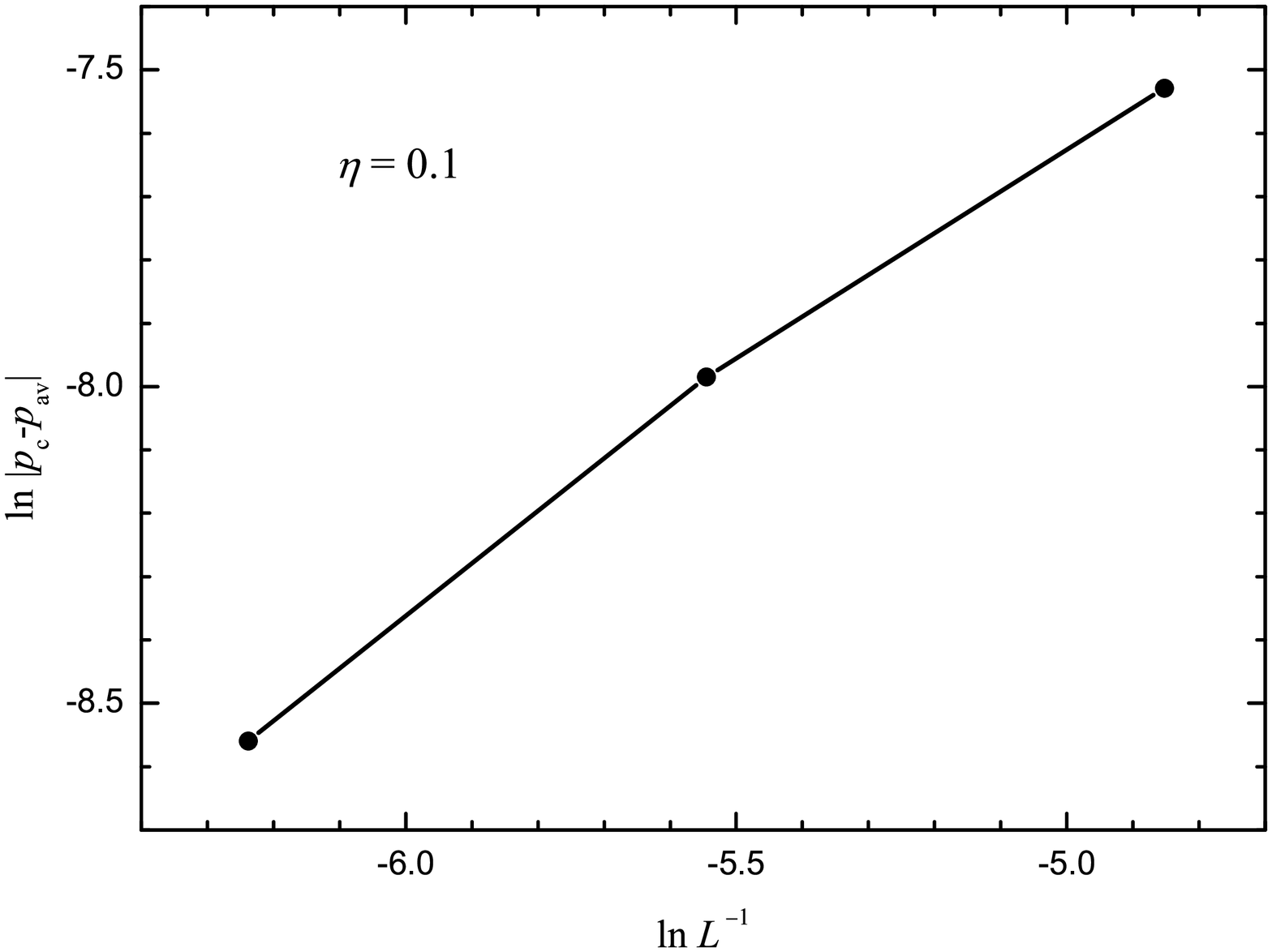}
\caption{\label{f4} The slope from the linear fitting of $\ln |p_c-p_{av}|$ versus $\ln L^{-1}$ gives the reciprocal of
$\nu$. Here, $p_c=0.530298$ for the model $\eta=0.1$.}
\end{figure}

As an example, the exponent $\nu=1.344(89)$ for $\eta=0.1$ is extracted as shown in FIG.~\ref{f4}. According to their
respective scaling relations, the values of $\beta$ and $\gamma$ for $\eta=0.1$ can be obtained in a similar way.
Finally, the values of exponents $\nu$, $\beta$, and $\gamma$ for our selected concentration of seeds from 0.05 to 0.5 are
summarized in Table~\ref{exps}. It can be seen that, there are minor differences in the values of each exponential for
different $\eta$ values, the values of exponents are in fact invariant, and they are respectively close to the
corresponding values of exponents in an ordinary two-dimensional site percolation model. The ratio between exponents
$(2\beta+\gamma)/\nu$ is therefore a constant, and it obeys the scaling law Eq.~(\ref{scalinglaw}) as it does in an usual
site percolation model.
\begin{table}[h]
  \centering
  \caption{The values of exponents $\nu$, $\beta$, $\gamma$, and rate $(2\beta+\gamma)/\nu$ under different concentration of seeds
  $\eta$.}\label{exps}
\begin{tabular}{r|c|c|c|c}
  \hline
    $\eta$ & $\nu$ & $\beta$ & $\gamma$ & $(2\beta+\gamma)/\nu$ \\ \hline
    0.05 & 1.35(17) & 0.142(18) & 2.39(30) & 1.98(50) \\
    0.1  & 1.344(89) & 0.138(11) & 2.38(16) & 1.98(27)\\
    0.2  & 1.359(40) & 0.141(7) & 2.41(7) & 1.98(12) \\
    0.3  & 1.353(85) & 0.139(10) & 2.39(15) & 1.97(25) \\
    0.4  & 1.356(79) & 0.140(10) & 2.40(14) & 1.98(23) \\
    0.5  & 1.353(83) & 0.140(10) & 2.39(15) & 1.97(24) \\ \hline
\end{tabular}
\end{table}

\section{Conclusion}
Given a number of seeds (except only one seed) on any lattice, percolation phase transition will inevitably happen while
clusters centered with these seeds grow up. Except the universal scaling law and various scaling exponents, the percolation
threshold is variable with non-uniform distribution of sites characterized by the concentration of seeds. Percolation
properties of a lattice depend on the non-uniform distribution of sites or bonds centered with multiple seeds, as well as
its structures. Any details (uniform or non-uniform structures) of geometrical distribution of sites or bonds may
contribute to the geometrical phase transition of a lattice. In general, smaller number of seeds for clusters growing
implies the earlier occurrence of percolation transition. It is expected that there exists a critical value of the
concentration of seeds, only below which, the abnormal finite-size scaling behaviours could happen.

The idea of multiple seeds can be extended to other correlated percolation models. This work make it possible to push the
application of percolation theory to wider fields, where percolation thresholds are expected to vary with non-uniform
population of sites or bonds while scaling exponents keep invariant.

Upon finishing this work, we became aware of a similar cluster model with initial seed concentration $\rho$ and an
additional parameter called growth probability $g$ reported by Roy and Santra recently~\citep{roy2013}. However, our model
(corresponding to their $g=1$) has not been discussed there. Except the difference in the model itself, the results of
their model, including the values of percolation thresholds around 0.593 for the selected seed concentration 0.05, 0.25,
and 0.50, and all the exponents are nearly the same for an ordinary percolation model.



\begin{thebibliography}{99}
\bibitem{stauffer} D. Stauffer and A. Aharony, {\em Introduction to Percolation Theory}, 2nd ed. (Taylor \& Francis, London, 1992).
\bibitem{hehre2003} W.J. Hehre, {\em A Guide to Molecular Mechanics and Quantum Chemical Calculations},
                    (Wavefunction, Inc., Irvine, California, 2003).
\bibitem{sahimi1994} M. Sahimi, {\em Applications of Percolation Theory} (Taylor \& Francis, Bristol, MA, 1994).
\bibitem{hunt2009} A. G. Hunt and R. Ewing, {\em Percolation Theory for Flow in Porous Media}, Lecture Notes in Physics 771, 2nd ed.,
                   (Springer, Berlin, 2009).
\bibitem{achlio2009} D. Achlioptas, R. M. D'Souza and J. Spencer, Science {\bf 323}, 1453 (2009).
\bibitem{lai2010} K. Lai, M. Nakamura, W. Kundhikanjana, M. Kawasaki, Y. Tokura, M. A. Kelly and Z.-X. Shen,
                  Science {\bf 329}, 190 (2010).
\bibitem{sergey2010} S. V. Buldyrev, R. Parshani, G. Paul, H. E. Stanley and S. Havlin,  Nature {\bf 464}, 1025 (2010).
\bibitem{costa2010} R. A. da Costa, S. N. Dorogovtsev, A. V. Goltsev and J. F. F. Mendes, Phys. Rev. Lett. {\bf 105}, 255701 (2010).
\bibitem{riordan2011} O. Riordan and L. Warnke, Science {\bf 333}, 322 (2011).
\bibitem{adler1991} J. Adler, Physica A {\bf 171}, 453 (1991).
\bibitem{paolo2004} P. D. Gregorio, A. Lawlor, P. Bradley and K.~A. Dawson, Phys. Rev. Lett. {\bf 93}, 025501 (2004).
\bibitem{balogh2009} J. Balogh, B. Bollob\'as, and R. Morris, Ann. Probab. {\bf 37}, 1329 (2009).
\bibitem{sausset2010} F. Sausset, C. Toninelli, G. Biroli, and G. Tarjus, J. Stat. Phys. {\bf 138}, 411 (2010).
\bibitem{fisher2006} C. Toninelli, G. Biroli, and D. S. Fisher, Phys. Rev. Lett. {\bf 96}, 035702 (2006).
\bibitem{toninelli2008} C. Toninelli and G. Biroli, J. Stat. Phys. {\bf 130}, 83 (2008).
\bibitem{jeng2008} M. Jeng and J. M. Schwarz, J. Stat. Phys. {\bf 131}, 575 (2008).
\bibitem{ghosh2014} A. Ghosh, E. Teomy and Y. Shokef, EPL {\bf 106}, 16003 (2014).
\bibitem{hinrichsen2000} H. Hinrichsen, Adv. Phys. {\bf 49}, 815 (2000).
\bibitem{odor2004} G. \'Odor, Rev. Mod. Phys. {\bf 76}, 663 (2004).
\bibitem{hinrichsen2006} H. Hinrichsen, Physica A {\bf 369}, 1 (2006).
\bibitem{henkel2008} M. Henkel, H. Hinrichsen, and S. L\"ubeck, {\em Non Equilibrium Phase Transitions, vol. 1: Absorbing
                     Phase Transitions} (Springer, 2008).
\bibitem{zhou2012} Z. Zhou, J. Yang, R. M. Ziff, and Y. Deng, Phys. Rev. E {\bf 86}, 021102 (2012).
\bibitem{lipowski2012} A. Lipowski, A. L. Ferreira, and J. Wendykier, Phys. Rev. E {\bf 86}, 041138 (2012).
\bibitem{landes2012} F. Landes, A. Rosso, and E. A. Jagla, Phys. Rev. E {\bf 86}, 041150 (2012).
\bibitem{wang2012} J. Wang, Z. Zhou, Q. Liu, T. M. Garoni, and Y. Deng, Phys. Rev. E {\bf 88}, 042102 (2012).
\bibitem{eden1961} M. Eden,  {\em Proc. 4th Berkeley Symp. on Mathematical Statistics and Probability, vol. 4,
                   ed. J. Neyman} (University of California Press, Berkeley,1961), p.~223.
\bibitem{jullien1985} R. Jullien and R. Botet, J. Phys. A {\bf 18}, 2279 (1985).
\bibitem{feng2008} X. Feng, Y. Deng and H. W. J. Bl\"{o}te, Phys. Rev. E {\bf 78}, 031136 (2008).
\bibitem{lee2008} M. J. Lee, Phys. Rev. E {\bf 78}, 031131 (2008).
\bibitem{yang2012} H. Yang, Phys. Rev. E {\bf 85}, 042106 (2012).
\bibitem{roy2013} B. Roy and S. B. Santra, Croat. Chem. Acta {\bf 86}, 495 (2013).

\end{thebibliography}
\end{document}